\documentclass{IEEEtran}
\usepackage{cite}
\usepackage{amsmath,amssymb,amsfonts}
\usepackage{algorithmic}
\usepackage{graphicx,color}
\usepackage{textcomp}
\usepackage{listings}
\usepackage{xcolor}
\usepackage[frozencache,cachedir=.]{minted}
\usepackage{hyperref}
\usepackage{bm}
\usepackage{rotating}
\usepackage[numbers]{natbib}
\def\BibTeX{{\rm B\kern-.05em{\sc i\kern-.025em b}\kern-.08em
    T\kern-.1667em\lower.7ex\hbox{E}\kern-.125emX}}
\AtBeginDocument{\definecolor{ojcolor}{cmyk}{0.93,0.59,0.15,0.02}}
\def\OJlogo{\vspace{-12pt}\includegraphics[height=24pt]{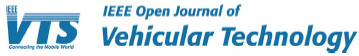}}

\begin{document}
%\receiveddate{XX Month, XXXX}
%\reviseddate{XX Month, XXXX}
%\accepteddate{XX Month, XXXX}
%\publisheddate{XX Month, XXXX}
%\currentdate{27 June, 2024}
%\doiinfo{OJVT.2024.0627000}

\title{FTA generation using GenAI with an Autonomy sensor Usecase}

\author{SNEHA SUDHIR SHETIYA,  IEEE SENIOR MEMBER, DIVYA GARIKAPATI , AUTONOMOUS SAFETY EXPERT, AND VEERAJA SOHONI AUTONOMOUS SAFETY EXPERT}

\maketitle

%\affil{National Institute of Standards and 
%Technology, Boulder, CO 80305 USA}
%\affil{Department of Physics, Colorado State University, Fort Collins, 
%CO 80523 USA}
%\corresp{CORRESPONDING AUTHOR: SNEHA SUDHIR SHETIYA (e-mail: sneha.shetiya@ieee.org).}
%\authornote{This work was supported by the Natural Sciences and Engineering Research Council (NSERC) of Canada.}
%\markboth{Preparation of Papers for IEEE OPEN JOURNALS}{Author \textit{et al.}}

\begin{abstract}
Functional safety forms an important aspect in the design of systems. Its emphasis on the automotive industry has evolved significantly over the years. Till date many methods have been developed to get appropriate FTA(Fault Tree analysis) for various scenarios and features pertaining to Autonomous Driving. This paper is an attempt to explore the scope of using Generative Artificial Intelligence(GenAI) in order to develop Fault Tree Analysis(FTA) with the use case of \textit{malfunction for the Lidar sensor} in mind. We explore various available open source Large Language Models(LLM) models and then dive deep into one of them to study its responses and provide our analysis. This paper successfully shows the possibility to train existing Large Language models through Prompt Engineering for fault tree analysis for any Autonomy usecase aided with PlantUML tool.
\end{abstract}

\begin{IEEEkeywords}
Generative Artificial Intelligence(GenAI), PlantUML(Unified Modeling Language), FTA(Fault Tree Analysis), FMEA(Failure Mode and Effects Analysis), FuSa(Functional Safety), SOTIF (Safety Of The Intended Functionality)
\end{IEEEkeywords}


\section{Introduction}

\IEEEPARstart{W}ith the advent of technology , safety has become a crucial aspect today. According to \cite{KAPITONOV2022556}, it is mandatory to ensure safety into the products put into circulation in any country. Safety is more so relevant today due to the amount on equipment installed on the vehicles. Particularly in the Autonomy usecase, we have innumerable sensors like Lidar, Radar, Camera and others necessary to enable the self-driving features of the vehicle in consideration. The two most important standards which come to the mind when we think about safety of Automotive is ISO 26262(Functional Safety of Road Vehicles)\cite{pathak2024iso} and PAS-21448(Safety Of The Intended Functionality)\cite{schnellbach2019development}.

\subsection{FuSa(Functional Safety) and SOTIF(Safety Of The Intended Functionality)}
Algorithms are developed that can make decisions in real-time by weighing the potential outcomes of different actions. These algorithms must be transparent and understandable to ensure public trust. It is necessary to establish transparent and clear policies for how autonomous vehicles should behave in different scenarios and ODD (Operational Design Domain)\cite{automated2020avsc}conditions. This can be then communicated to the public to build trust and acceptance. One approach can be to combine FuSa and SOTIF principles to create a multi-layered safety approach that ensures both the reliability of the vehicle's systems and the safety of its intended functionalities. Data from real-world driving and simulations can be used to continually improve the decision-making algorithms, ensuring they remain effective as new ethical challenges and scenarios emerge. 
For FuSa and SOTIF to be merged and safety use cases to be created, the GenAI based approaches can be leveraged to come up with different analysis at different stages of FuSa/SOTIF HW/SW development.

\subsection{Gen AI Introduction and types}
\begin{figure}
\centerline{\includegraphics[width=3.5in]{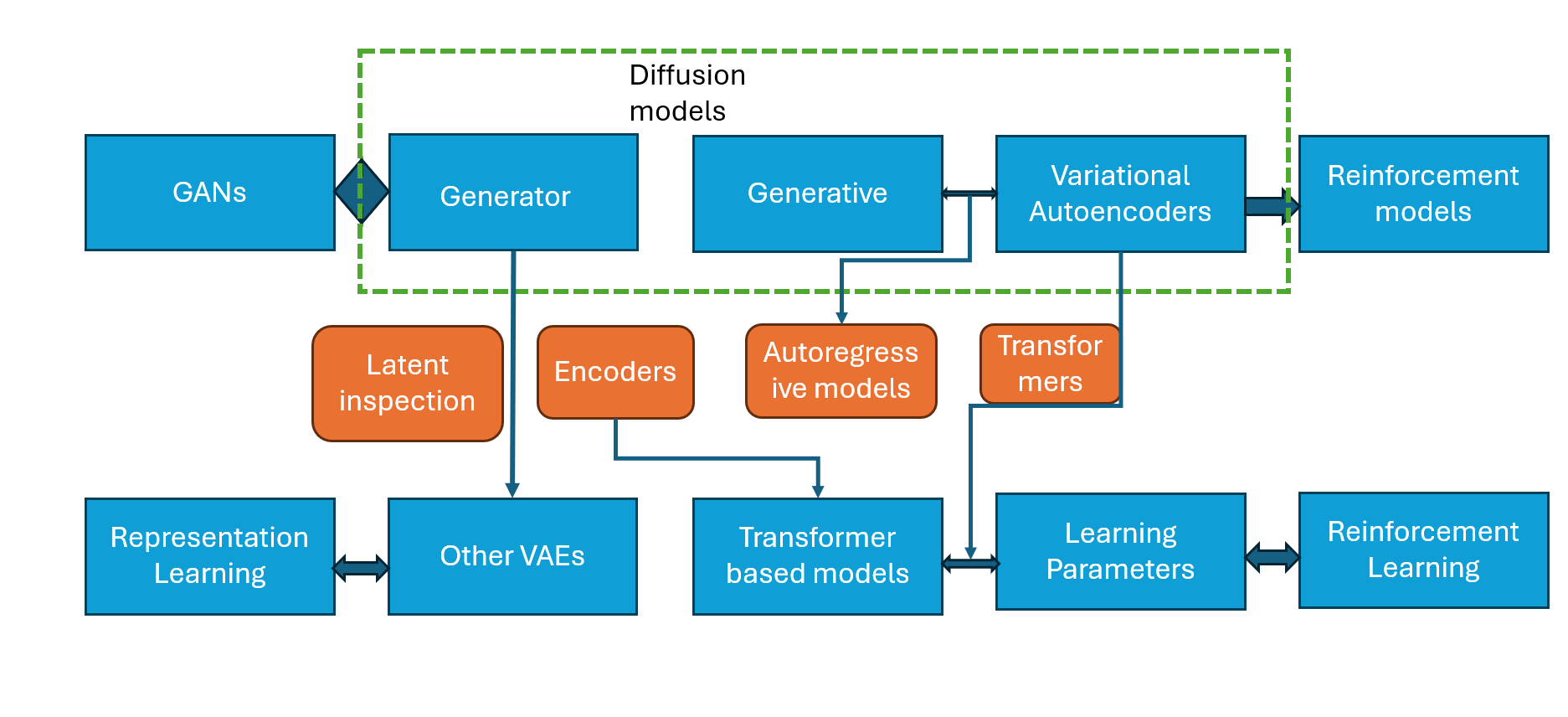}}
\caption{Common models in Generative AI domain}
\label{Common models in Generative AI domain}
\end{figure}

Generative Artificial Intelligence(GenAI) is a concept that has evolved from Artificial Intelligence\cite{10463265} . Earlier there was the advent of neural networks\cite{gurney2018introduction} which were built from basic mathematical principles of backpropagation [\cite{rojas1996backpropagation}]. Overtime these neural networks were used to train Machine Learning models for prediction algorithms. Multiple use cases of this were to study crop withering[\cite{nyarukowa2016swapdt}], object detection for basic self driving features[\cite{simhambhatla2019self}] and so on. Eventually Natural Language Processing gained prominence and paved the way for the advent of Large Language Models\cite{chowdhary2020natural}. The popular examples for these models today are OpenAI's ChatGPT-4 and Google's Gemini. Its common place now to utilize these technologies to leverage code optimizations and do other redundant work like framing emails and generating AI images.
GenAI in a nutshell can be defined as technology focused on creating new data often resembling existing data. Some common types of this technology are elaborated below\ref{Common models in Generative AI domain}:

1. Generative Adversarial Networks(GANs)
    This has two neural networks in its design: Generator and a Discriminator. Generator's job is to create new data and the discriminator, as the name suggests tries to distinguish real data from Generator's creation
    The output of this combination will give the desired creative result.
    This is majorly used in creating product designs, composing music and developing novel materials.

2. Variational Autoencoders (VAEs)
    VAEs also work with data. It used encoding/decoding techniques to generate new data. It has a neural network which compresses the training data when encoding and then decodes it back to reconstruct it.VAE introduces randomness when decoding. This helps to create variable outputs which are similar to the input data.
    This is majorly used to create variations in existing artwork, generate realistic 3D models and anomaly detection.

3. Autoregressive models
    These ML models generate data sequentially, predicting the next element based on the previously generated elements. This can be text, code or any other sequential data type.
    This is majorly used in creative writing, chatbot responses and music composition.

4. Transformer-based models
    These are another set of powerful neural network based models which deal with sequential data. These are used for generative tasks by conditioning the model on a specific point or a theme to generate new, relevant data
    This is majorly used in generating creative text formats like poems and scripts, translating languages and summarizing text content.

5. Reinforcement learning for Generative Tasks
    Here a reinforcement learning agent interacts with the surrounding environment  and receives good feedback for generating desired types of data. Over time, the agent learns to create increasingly better outputs based on the reward feedback.
    This has major applications in robotics, design game levels and autonomous systems.

\subsection{HAZOP(FMEA/FTA analysis)}
Hazard and Operability Study originally developed as an engineering tool utilized in the process industries to prevent occurrence of industrial hazards and plan mitigation strategies\cite{10124508}. Overtime, its applications increased in various industries including identifying risk and hazards in engineering systems. The focus of HAZOP is mainly to avoid workplace hazards. But we will be leveraging the contents given in the HAZOP process to effectively consider a usecase in Autonomous driving and perform safety analysis for the same.
Failure Mode and Effects Analysis (FMEA) \cite{European_Standards} is a process that identifies potential failures with assets and other areas of business. The benefits of utilizing FMEA include reducing potential failures, saving lives, and lowering excessive costs. FMEA analysis is out of scope for the current paper and will be elaborated in the next research paper.

\subsubsection{Benefits of FTA in Automotive}
\textbf{Proactive Risk Identification}: FTA helps pinpoint potential failures before they occur, preventing costly recalls and reputational damage. \\
\textbf{Improved Safety}: By identifying hazards and their root causes, OEMs can implement measures to enhance vehicle safety. \\
Enhanced Reliability: FTA contributes to designing and manufacturing vehicles that are less likely to break down, leading to increased customer satisfaction. \\
\textbf{Regulatory Compliance}: FTA can help ensure that vehicles meet safety standards and regulations, avoiding fines and legal issues.\\
\textbf{Cost Reduction}: Identifying potential failures early can save money by preventing costly repairs and downtime. \\
\textbf{Optimized Design and Manufacturing}: FTA can inform design decisions and manufacturing processes, leading to more reliable and efficient vehicles.\\

Fault Tree Analysis is a top-down approach originally developed in Bell laboratories by H. Watson and A. Mearns for the Air Force in 1962\cite{10124508}. This concept was later adopted by Boeing, and today, it is widely used in the aerospace, automobile, chemical, nuclear, and software industries, especially for reliability and safety events. 

FTA (within HAZOP context)\cite{HarnessingAI}
Definition: Within a HAZOP framework, FTA is used to provide a structured, graphical representation of the relationships between different failures and their potential causes. It can help in understanding the root causes of hazards identified in the HAZOP study and in evaluating the probability of different combinations of failures leading to an undesirable event.
Application: FTA can be employed after a HAZOP study to analyze complex interactions and dependencies between different failures, helping to identify critical points of failure and potential mitigation strategies.

\begin{figure}
\centerline{\includegraphics[width=3.5in]{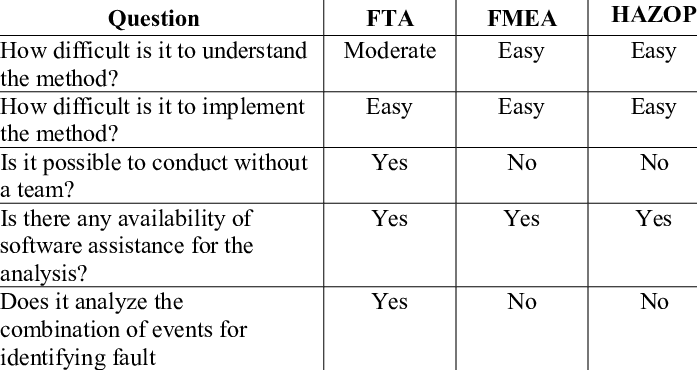}}
\caption{Comparisons between HAZOP/FMEA/FTA \cite{basnet2018review}
\label{A-general-comparison-between-FTA-FMEA-and-HAZOP-12}}
\end{figure}

Table\ref{A-general-comparison-between-FTA-FMEA-and-HAZOP-12}shows a list of basic questions one can ask to determine the work involved and its dependencies when it comes to FTA, FMEA and HAZOP.

\section{Existing Research}
Initial methods for effective FTA analysis have been illustrated in \cite{campean2008function}. This standard describes in detail the required risk analysis and evaluation procedure and safety measures, while appropriately reflecting on the system design definition. Therefore, much research has been done on the risk analysis procedure, wherein the design information is mostly extracted from physical components of similar systems already in operation, the information traced back to obtain constituent functions, and then methods of identifying risk sources are studied.  The paper has good conceptual design work done. Eventually its the job of safety engineers to leverage this to perform the fault tree analysis and risk analysis of the system in consideration.

Automotive Airbag system is the first non-classical feature of modern cars which complements the safety systems that protect the occupants of the vehicle.Airbags are majorly to protect in a crash situation, however may result in adverse effects if inadvertently deployed \cite{swarup2014safety}.This paper performs FTA analysis on the airbag system to identify the various faults that can lead to a top level undesired event, leading to an accident. This helps in getting early maturity of the airbag design in aspects of safety.

Coming to GenAI, \cite{silvis2024sart} focuses on the safety analysis of radiation therapy(RT), an effective technique to treat cancer. It involves complex machines, human operators and software which make it safety critical. Although the functionality is exclusive to Automotive, it has similar issues which are observed in safety analysis of Automotive vehicles: 
varying FMEA and FTA across RT Organizations. Thus i-SART , a GenAI based algorithms developed using NLP(Natural Language Processing) techniques is developed to promote global implementation of safe RT practices.

A recent research \cite{gimpel2023unlocking} is a white paper which talks about unlocking the power of generative AI models and systems such as GPT-4 and ChatGPT as a publicly accessible system to improve efficiency in teaching and learning concepts. This is an interesting thought process as in our current research our major focus is to leverage these models for technical analysis which are crucial to the design of Autonomous systems in vehicles.There is a goal in the Industry to explore on the concrete application possibilities of these ChatGPT type models. This paper is an attempt to make an increment towards the achievement of that goal.

\section{Need for Gen AI in FTA analysis}

The entire  FTA process currently involves a lot of manual analysis and the process can't be completed without taking help from multiple stakeholders. It is heavily based on the documentation and recording of different failure modes in a natural language. 

As the complexity of the automotive SoC(System On Chip) grows, there will be a need to digitize these FTAs. Digitization of FTA is for processing and analyzing the FTA data to understand the potential hardware/software risks. To improve the quality of FTA, there will be a growing need to use the co-pilot who is trained with the prior/parent FTA data which can be further enhanced by the co-pilot\cite{lu2022safety} by regenerating failure modes based on this empirical data. In the context of AI, a co-pilot is an AI-powered virtual assistant designed to work alongside users, providing guidance, assistance, and even completing tasks on their behalf. It leverages large language models (LLMs) to understand and process information, and can be used in various domains and industries. For example, AI can analyze large datasets (e.g., historical failure data, and maintenance logs) to identify common failure modes and their causes. This can help in quickly generating an initial list of potential failures.  AI algorithms can automate the probability calculation of various complex events by estimating the probability of occurrence of simple events.

Major contribution of the paper:
1. The paper discusses on various Large Language Models available in the Industry under Generative AI domain.
2. Initial tests are proved on an open source LLM model which shows promising results.
3. Experiments are conducted for various positive and negative scenarios for FTA generation using a GenAI model in the form of UML(Unified Modeling Language) code.
4. PlantUML is explored in its entirety to generate FTA diagrams.
5. Successful experiments open up scope for future work.

\section{Standardized shapes for FTA}
Several standards documents and technical papers related to Fault Tree Analysis (FTA) have been published, outlining standard shapes and methodologies used in this analysis.\cite{backstrom2016effective} Here are some key references:
NRC NUREG–0492: This standard\cite{tankaas} is specifically tailored for the nuclear power industry and outlines methodologies for FTA. It has been revised for aerospace applications by NASA.
SAE ARP4761: This is a standard\cite{leveson2014comparison} for civil aerospace, providing guidelines for performing FTA in the context of aviation safety.
MIL–HDBK–338: This military handbook\cite{moon1966use} addresses FTA methodologies applicable to military systems, ensuring reliability and safety in defense applications.
IEC 61025: This international standard\cite{zornoza2018model} is designed for cross-industry use and has been adopted as European Norm EN 61025, providing a comprehensive framework for conducting FTA across various sectors.
These documents collectively provide a robust foundation for understanding and implementing Fault Tree Analysis, including the standard shapes and logic used in constructing fault trees.
For more specific standards related to transit and mobility data, the FTA Report No. 267\cite{FTA_Report} discusses interoperability standards and specifications, which may also include graphical representations relevant to FTA methodologies in public transportation contexts.

\subsection{Most commonly used standard shapes in FTA}
\begin{figure}
\centerline{\includegraphics[width=3.5in]{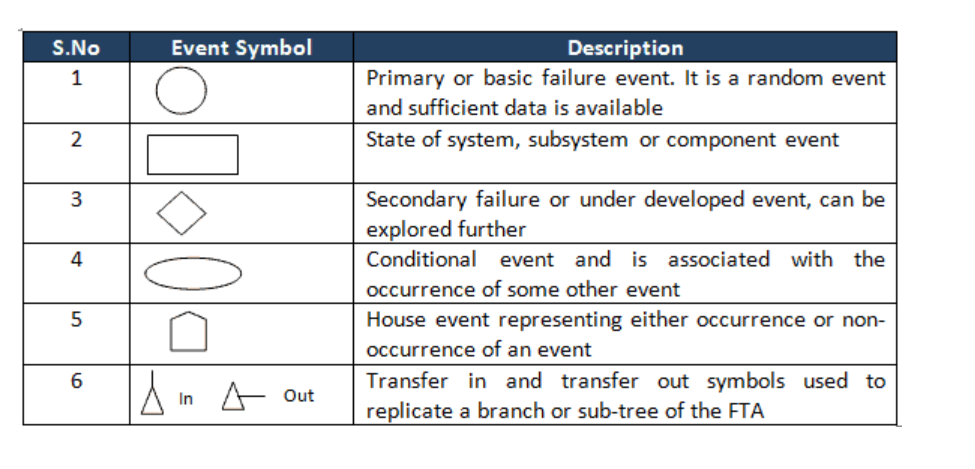}}
\caption{Standard symbols used for FTA \cite{SixSigma}
\label{FTA_symbols}}
\end{figure}

In Fault Tree Analysis (FTA), several standard shapes and symbols are commonly used to represent events and logical relationships\ref{FTA_symbols}. Here are the key shapes: \\
1. \textbf{Event Symbols}
Basic Event: Represents a failure or error in a system component (e.g., a switch stuck in the open position).\\
2. \textbf{External Event:} Indicates an event that is normally expected to occur due to external factors but is not a fault.\\
3. \textbf{Undeveloped Event:} Represents an event with insufficient information or one that is deemed inconsequential.\\
4. \textbf{Conditioning Event:} Reflects conditions that affect the logic gates (e.g., mode of operation).

\subsubsection{Gate Symbols}
OR Gate: The output occurs if any input occurs.
AND Gate: The output occurs only if all inputs occur.
Exclusive OR Gate: The output occurs if exactly one input occurs.
Priority AND Gate: The output occurs if inputs occur in a specified sequence.
Inhibit Gate: The output occurs if the input occurs under a specific enabling condition.

\subsubsection{Transfer Symbols}
Transfer Symbols: Used to connect the inputs and outputs of related fault trees, such as linking a subsystem's fault tree to its overall system fault tree.
These symbols are essential for visually representing the relationships and events in an FTA, facilitating a clear understanding of system failures and their causes.

\subsection{Benefits of using standardized shapes in FTA}
Using standardized shapes in Fault Tree Analysis (FTA) provides several key benefits:

\subsubsection{Improved Clarity and Understanding}
The use of standardized event, gate, and transfer symbols creates a common visual language that makes fault trees easier to interpret and communicate. This clarity facilitates collaboration among diverse stakeholders like engineers, maintenance teams, and management.
\subsubsection{Consistency and Comparability}
Adhering to standard shapes enables consistency in FTA modeling across an organization. This allows for better comparison of fault trees, identification of common failure patterns, and benchmarking of results.
\subsubsection{Alignment with Industry Best Practices}
Major standards like NRC NUREG-0492, SAE ARP4761, and IEC 61025 specify the standard shapes to be used in FTA. Employing these shapes ensures alignment with industry best practices and facilitates compliance with regulations.
\subsubsection{Ease of Interpretation}
The symbolic representations of events, gates, and transfers intuitively convey the relationships and logic in a fault tree. This makes the diagrams easier to interpret, especially for complex systems.
\subsubsection{Compatibility with FTA Software}
FTA software tools are designed to work with the standard shapes. Using these shapes ensures compatibility and enables the use of automated analysis features provided by the software.

In summary, standardized shapes are essential for creating clear, consistent, and industry-aligned fault trees that can be readily interpreted and analyzed. They form a critical component of effective FTA practice.

\subsection{Leverage these standardized shapes to help train individuals or models to do an FTA on a component described}
To leverage standardized shapes in Fault Tree Analysis (FTA) for training purposes, the following strategies can be employed:

1. Structured Training Modules
Develop training modules that focus on the specific standardized shapes used in FTA, such as basic events, gates (AND, OR, etc.), and transfer symbols. Each module can include:
Visual Aids: Use diagrams to illustrate each shape and its meaning.
Examples: Provide real-world examples of fault trees that utilize these shapes effectively.

2. Interactive Learning Tools
Utilize software tools that allow trainees to create fault trees using standardized shapes. This hands-on approach can include:
Simulation Exercises: Create scenarios where trainees must identify potential failures and construct a fault tree using the correct shapes.
Feedback Mechanisms: Implement systems for providing immediate feedback on the accuracy of the fault tree diagrams created by trainees.

3. Collaborative Workshops
Organize workshops that encourage collaboration among participants from different disciplines. This can enhance understanding through:
Group Activities: Facilitate group exercises where participants work together to analyze a component and construct a fault tree.
Cross-Disciplinary Perspectives: Involve engineers, maintenance personnel, and safety experts to provide diverse insights on the component being analyzed.

4. Case Studies and Best Practices
Incorporate case studies that highlight successful applications of FTA in various industries. Focus on:
Lessons Learned: Discuss how standardized shapes helped identify critical failures and improve system reliability.
Regulatory Compliance: Emphasize how using standardized shapes aligns with industry regulations and best practices, enhancing safety and compliance efforts.

5. Continuous Improvement and Updates
Encourage ongoing learning by:
Regularly Updating Training Materials: As new standards or shapes emerge, ensure training content reflects the most current practices.
Feedback Loops: Collect feedback from trainees to continuously improve the training process and materials.
By implementing these strategies, organizations can effectively train personnel in FTA using standardized shapes, ultimately leading to better risk management and system reliability.

\section{Using LLMs to produce basic FTA models}
We used several publicly available LLMs like OpenAI's ChatGPT\cite{ChatGPT}, Google's Gemini\cite{Gemini} and Perplexity\cite{Perplexity} to train them to generate desired Fault Tree analysis through Prompt Engineering. Following are the results:

1. Gemini: When we requested Gemini to generate an FTA for the lidar sensor keeping in mind Autonomous Driving, it gave us links\cite{bing}\cite{wiley} to existing research which do FTA analysis. We went a step ahead and requested it to produce by itself. It then gave a theoretical explanation which was a standard FTA diagram with no reference to Lidar and also revealed its incapability to generate images.

\begin{figure}
\centerline{\includegraphics[width=3.5in]{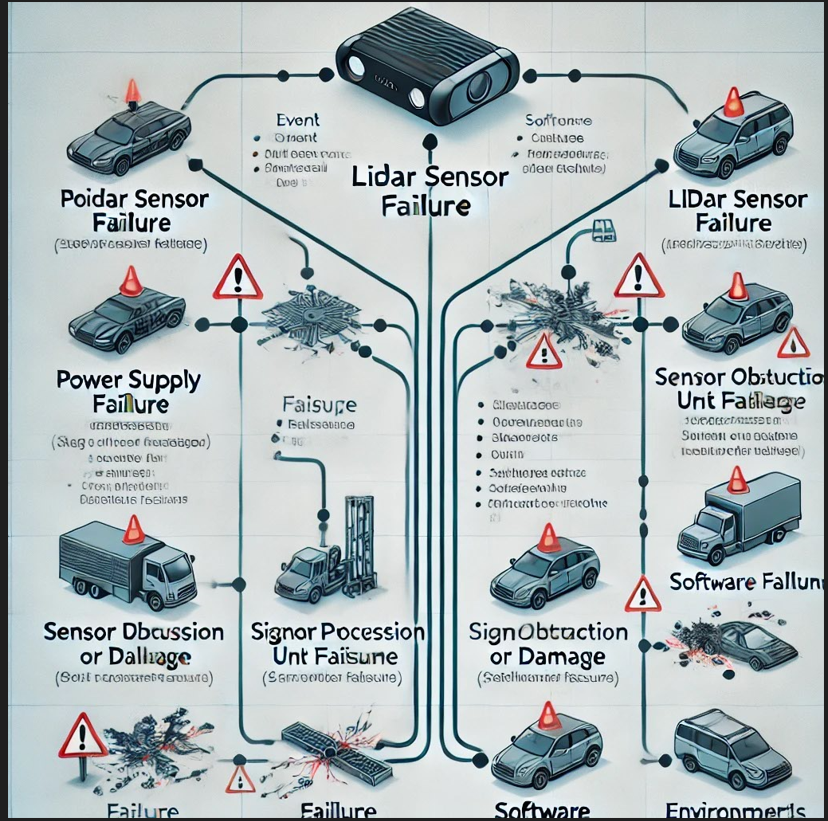}}
\caption{ChatGPT generated initial FTA}
\label{ChatGPT generated initial FTA}
\end{figure}

2. ChatGPT: This model was smarter in its approach than Gemini. Initially when we requested it the same question like Gemini to generate an FTA for the lidar sensor, it gave theoretical approach with detailed analysis to do the desired FTA. It also requested us if we need it in graphical format. Upon confirming , figure \ref{ChatGPT generated initial FTA} was generated . However this did not suit our needs as we needed a flowchart with standard gates as mentioned in earlier sections.

From the figure\ref{ChatGPT generated initial FTA}e, we can observe that the image is not readable. The model has tried its best to represent the hardware and its subsequent fault analysis. However it cannot pass as valid FTA.

\begin{figure}
\centerline{\includegraphics[width=3.5in]{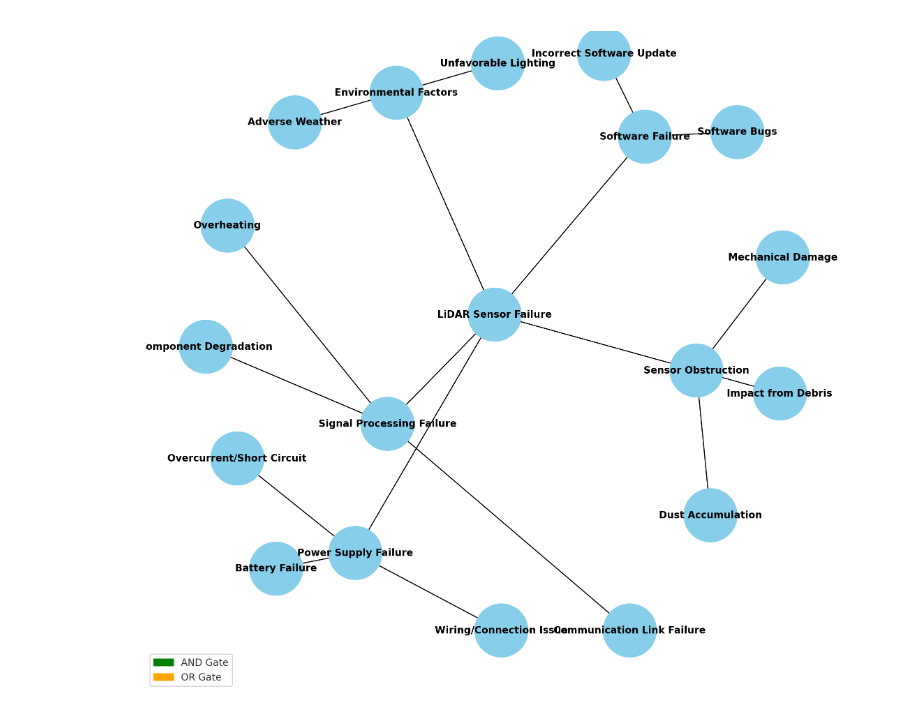}}
\caption{ChatGPT generated final FTA}
\label{ChatGPT generated final FTA}
\end{figure}

We fine tuned our prompt to get a figure with gates. Figure \ref{ChatGPT generated final FTA} was generated that showed the mapping, but failed to produce the desired result.

3. Perplexity: This model seemed to be promising as we were able to make great progress in training it. Repeating the initial prompt to generate an FTA for lidar sensor in autonomy, figure \ref{Perplexity generated initial FTA} was generated:

\begin{figure}
\centerline{\includegraphics[width=3.5in]{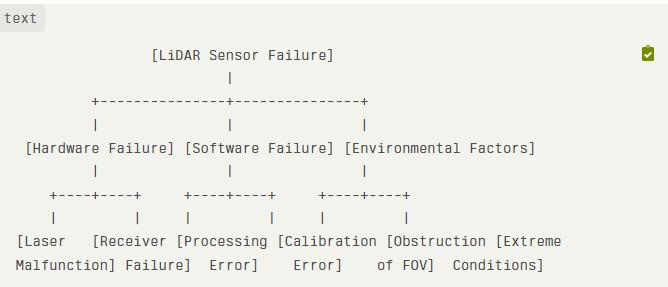}}
\caption{Perplexity generated initial FTA}
\label{Perplexity generated initial FTA}
\end{figure}

\begin{figure}
\centerline{\includegraphics[width=3.5in]{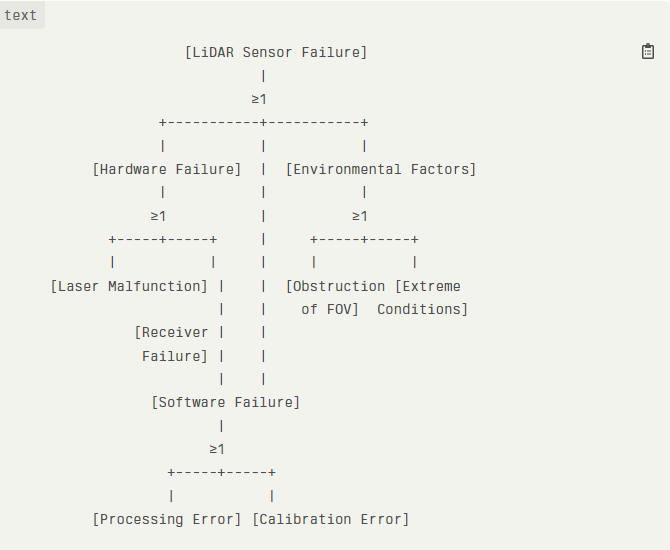}}
\caption{Perplexity generated second FTA}
\label{Perplexity generated second FTA}
\end{figure}

Next we asked it to represent figure\ref{Perplexity generated initial FTA} it with gates to obtain the figure\ref{Perplexity generated second FTA}

In this diagram\ref{Perplexity generated second FTA}:
$\geq $represents an OR gate. This symbol means that if any one (or more) of the input events occurs, the output event will occur.

\begin{figure}
\centerline{\includegraphics[width=3.5in]{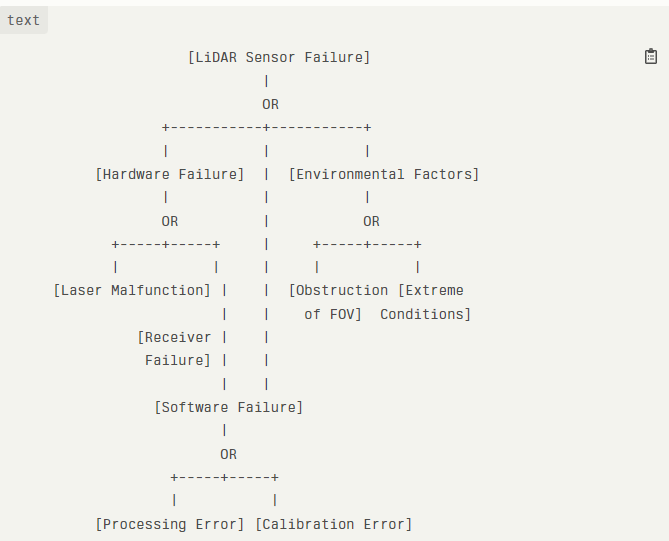}}
\caption{Perplexity generated final FTA}
\label{Perplexity generated final FTA}
\end{figure}

Rectangle: Represents an event
$\geq $: OR gate
This diagram now uses the standard symbols for fault tree analysis, making it more aligned with professional FTA practices. 
Next as we have seen in standard shapes, we do not use above signs but actual AND/OR gates to represent the transitions. Thus we requested it again to generate with AND/OR gates and got the response as shown in figure \ref{Perplexity generated final FTA}. It replaced it with the symbols AND/OR but did not get gates into its flowchart. We could only get so far with perplexity.

This is where we stopped with the above three models.

\subsection{Major challenges faced in this experiment} 
1. We realized that although we can use these standardized LLMs to get FTAs, we cannot get the standardized flowchart shapes.
2. There is a need to build an example LLM model that recognizes these shapes.
3. Need to evaluate possibility of different modeling techniques and sync them with Perplexity to get FTAs.

In order to overcome above challenges, we took the aid of PlantUML\cite{lund2022towards}.
PlantUML is a powerful tool that allows you to create various types of diagrams using a simple text-based syntax. This means you can define diagrams using plain text, which can be easily integrated into your documentation or source code.
The ease with which this can be integrated in our VS code environment was another major factor for us to choose this line of research for next experiments.

\section{Utilise code generation capabilities of GenAI models to generate FTAs}

\begin{figure}
\centerline{\includegraphics[width=3.5in]{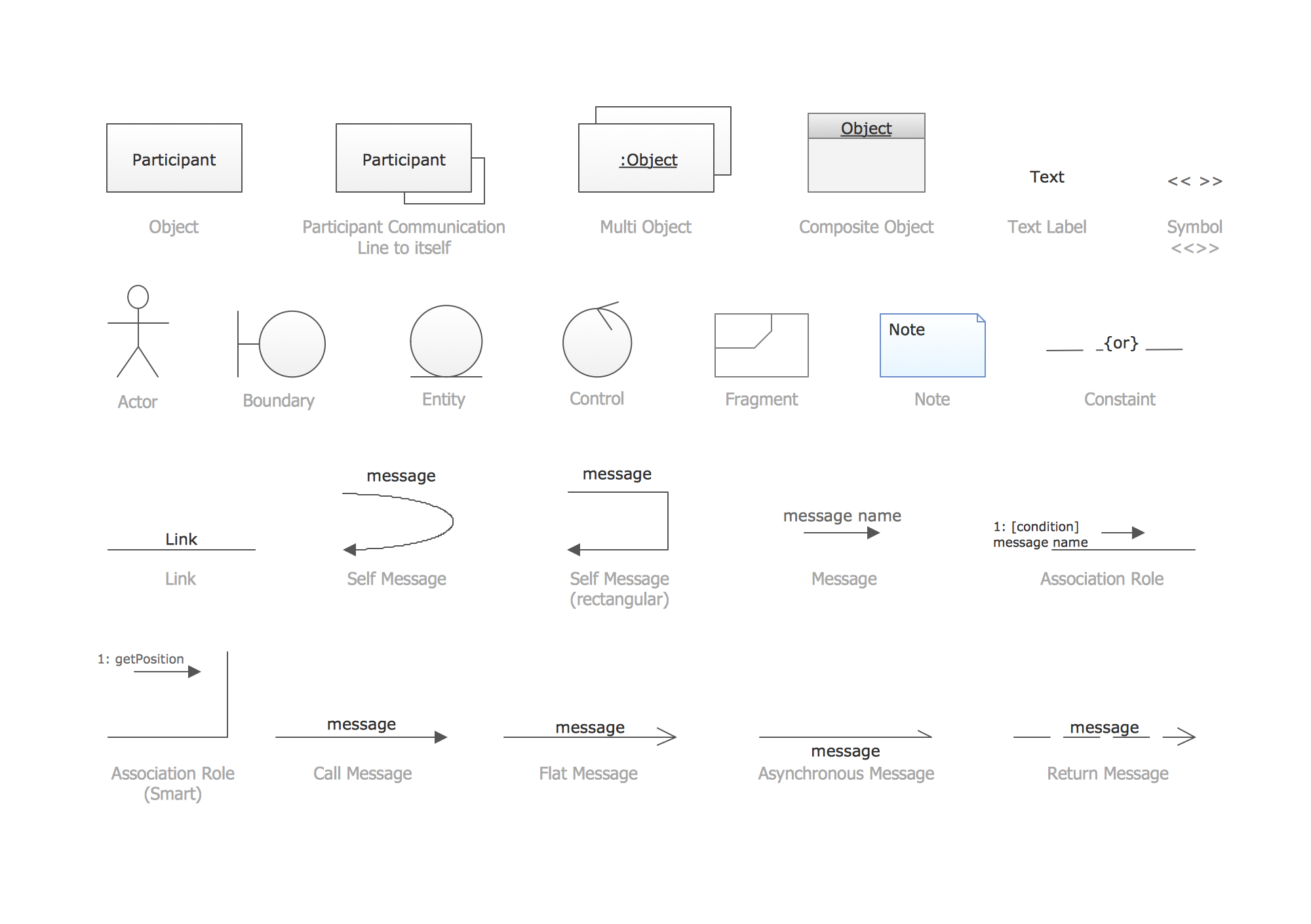}}
\caption{Basic symbols for PlantUML  \cite{HowToUML}
\label{Basic symbols for PlantUML}}
\end{figure}

Like FTA, UML also has standard shapes illustrated in table \ref{Basic symbols for PlantUML}

\subsection{PlantUML Initial FTA}

The series of prompts provided to Perplexity to generate our first PlantUML diagram  is illustrated in Table\ref{tab2}:

\begin{table}
\caption{Prompt Sequence for FTA}
\label{tab2}
\setlength{\tabcolsep}{3pt}
\begin{tabular}{|p{25pt}|p{100pt}|p{80pt}|}
\hline
Sl No.& 
Prompt& 
Output \\
\hline
1 & 
Generate FTA for Lidar sensor in Autonomy& 
Theory with directions to generate FTA \\
\hline
2 &
Please generate it with gates as a Fault tree analysis diagram&
Figure \ref{Perplexity generated second FTA}\\
\hline
3 &
Can you convert above FTA diagram(referring to \ref{lst:PlantUMLInitialFTA}) to UML code to generate a PlantUML diagram?&
Generated necessary PlantUML code\\

\end{tabular}
\end{table}

\begin{figure*}[htbp]
    \centering
    \begin{turn}{90}
    \includegraphics[width=1.3\textwidth,height=1.0in]{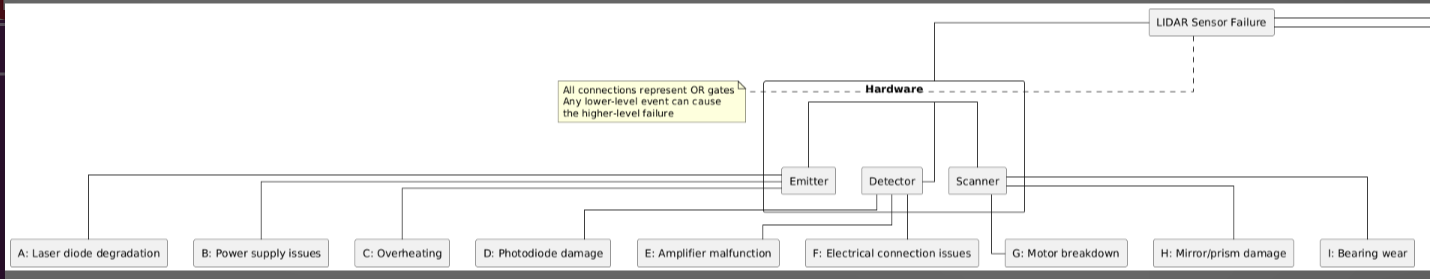}
    \end{turn}
    \caption{Plant UML generated FTA for autonomy left flowchart}
    \label{fig:Plant UML generated FTA for autonomy left flowchart}
\end{figure*}

\begin{figure*}[htbp]
    \centering
    \begin{turn}{90}
    \includegraphics[width=1.3\textwidth,height=1.0in]{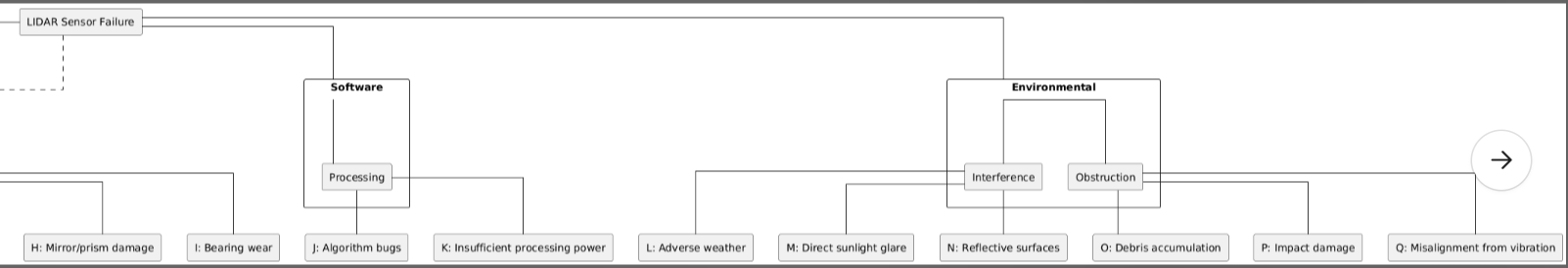}
    \end{turn}
    \caption{Plant UML generated FTA for autonomy right flowchart}
    \label{fig:Plant UML generated FTA for autonomy right flowchart}
\end{figure*}

Code block\ref{lst:PlantUMLInitialFTA}) was generated and the resultant figure is figure\ref{fig:Plant UML generated FTA for autonomy left flowchart} and \ref{fig:Plant UML generated FTA for autonomy right flowchart}. These have been split into two diagrams for better viewability as the diagrams were getting blurry when included in paper as a whole.
\lstset{
  language=Python,         % Specify the language for syntax highlighting
  breaklines=true,       % Enable line breaking
  breakatwhitespace=true,% Only break at whitespace
  basicstyle=\ttfamily,    % Code font style
  keywordstyle=\color{blue}, % Color of keywords
  commentstyle=\color{gray}, % Style for comments
  stringstyle=\color{red},   % Style for strings
  showstringspaces=false,   % Prevents showing spaces as special symbols
  %numbers=left,            % Line numbers on the left
  %numberstyle=\tiny\color{gray}, % Style for line numbers
  frame=single,            % Puts a frame around the code
}

%\begin{table}[h]

  %  \centering
  %  \caption{\textbf{Code to generate Perplexity generated FTA for first %time using PlantUML}}
  %  \label{lst:PlantUMLInitialFTA}
%\begin{lstlisting}[\caption{\textbf{Code to generate Perplexity %generated FTA for first time using %PlantUML}},\label{lst:PlantUMLInitialFTA}]
\begin{lstlisting}[caption={\textbf{Code to generate Perplexity generated FTA for first time using PlantUML}}, label=lst:PlantUMLInitialFTA]
    @startuml LIDAR Sensor Failure FTA
    skinparam packageStyle rectangle
    skinparam linetype ortho
    
    rectangle "LIDAR Sensor Failure" as TopEvent
    
    package "Hardware" {
      rectangle "Emitter" as Emitter
      rectangle "Detector" as Detector
      rectangle "Scanner" as Scanner
    }
    
    package "Software" {
      rectangle "Processing" as Processing
    }
    
    package "Environmental" {
      rectangle "Interference" as Interference
      rectangle "Obstruction" as Obstruction
    }
    
    TopEvent -- Hardware
    TopEvent -- Software
    TopEvent -- Environmental
    
    Hardware -- Emitter
    Hardware -- Detector
    Hardware -- Scanner
    
    Software -- Processing
    
    Environmental -- Interference
    Environmental -- Obstruction
    
    rectangle "A: Laser diode degradation" as A
    rectangle "B: Power supply issues" as B
    rectangle "C: Overheating" as C
    Emitter -- A
    Emitter -- B
    Emitter -- C
    
    rectangle "D: Photodiode damage" as D
    rectangle "E: Amplifier malfunction" as E
    rectangle "F: Electrical connection issues" as F
    Detector -- D
    Detector -- E
    Detector -- F
    
    rectangle "G: Motor breakdown" as G
    rectangle "H: Mirror/prism damage" as H
    rectangle "I: Bearing wear" as I
    Scanner -- G
    Scanner -- H
    Scanner -- I
    
    rectangle "J: Algorithm bugs" as J
    rectangle "K: Insufficient processing power" as K
    Processing -- J
    Processing -- K
    
    rectangle "L: Adverse weather" as L
    rectangle "M: Direct sunlight glare" as M
    rectangle "N: Reflective surfaces" as N
    Interference -- L
    Interference -- M
    Interference -- N
    
    rectangle "O: Debris accumulation" as O
    rectangle "P: Impact damage" as P
    rectangle "Q: Misalignment from vibration" as Q
    Obstruction -- O
    Obstruction -- P
    Obstruction -- Q
    
    note bottom of TopEvent
      All connections represent OR gates
      Any lower-level event can cause
      the higher-level failure
    end note
    
    @enduml


    \end{lstlisting}
%\end{table} % Use figure or listing environment

%\begin{table*}[htbp]
%\centering
%\begin{tabular}{c|c}
%\hline
%Column 1 & Column 2 \\
%\hline
%\multicolumn{2}{c}{\includegraphics[width=\linewidth, height=2in]{Initial_FTA.PNG}} \\
%\hline
%\end{tabular}
%\caption{Plant UML generated FTA for autonomy}
%\label{Plant UML generated FTA for autonomy}
%\end{table*}

%\begin{figure*}[h!]
%\centering
%\begin{tabular}{c|c}
%\hline
%Column 1 & Column 2 \\
%\hline
%\begin{turn}{90}
%\includegraphics[width=3\linewidth]{Initial_FTA.PNG}
%\hline
%\end{turn}
%\end{tabular}
%\caption{Plant UML generated FTA for autonomy}
%\label{Plant UML generated FTA for autonomy}
%\end{figure}

The code is straight forward with Top Events , nodes and connections. Its interesting to understand how Perplexity provides its analysis of the output it gives. This is in further sections.\ref{subsec:Error}

\section{Training and Fine Tuning}
\subsection{Evaluate potential PlantUML code generators}
Before fixing upon our preferred way to get FTA, we evaluated potential tools to best fit our usecase.
\subsubsection{TopEvent FTA}
This tool\cite{FTA_Tool} was available open source and an Industry standard used in automotive FTA designs. Based upon our analysis, this tool was found to be specific to particular ECU(Electronic Control Unit) use cases and not fit to be used with our model.

\subsubsection{Mermaid}
This tool\cite{Mermaid_Tool} is a popular to generate sequence diagrams in software applications. However as we needed more top-down analysis for our FTA and less sequences , we decided not to use Mermaid.

\subsubsection{yUML}
This is a web-based tool\cite{yUML_Tool} that allows you to create UML diagrams using a simple text syntax. However we found we could use the simple tax which yUML uses but have it in-built as an extension in our VS code.

\subsubsection{Nomnoml and Kroki}
These tools\cite{nomnoml_Tool}\cite{kroki_Tool} are good for class diagrams and Kroki in particular is an API which uses backend to generate diagrams.
As our usecase was not necessary software based, we decided not to use these approaches.

We decided to use syntax similar to yUML but plug it in as an extension with VS code on our system to generate and verify UML diagrams.

\subsection{PlantUML modified FTA to fix errors}
\label{subsec:Error}
\subsubsection{Prompt 1:give a Plant UML code to generate an FTA for lidar sensor in for level 4 autonomy in all weather conditions}
Code generated: \ref{lst:PlantUMLmodifiedFTA}\\

\begin{lstlisting}
[caption{\textbf{Code to generate Perplexity generated FTA after all modifications using PlantUML}}, label=lst:PlantUMLmodifiedFTA]
@startuml LIDAR Sensor Failure FTA

skinparam packageStyle rectangle
skinparam linetype ortho

rectangle "LIDAR Sensor Failure" as TopEvent

package "Hardware" {
  rectangle "Emitter" as Emitter
  rectangle "Detector" as Detector
  rectangle "Scanner" as Scanner
}

package "Software" {
  rectangle "Processing" as Processing
}

package "Environmental" {
  rectangle "Interference" as Interference
  rectangle "Obstruction" as Obstruction
}

TopEvent -- Hardware
TopEvent -- Software
TopEvent -- Environmental

Hardware -- Emitter
Hardware -- Detector
Hardware -- Scanner

Software -- Processing

Environmental -- Interference
Environmental -- Obstruction

rectangle "A: Laser diode degradation" as A
rectangle "B: Power supply issues" as B
rectangle "C: Overheating" as C
Emitter -- A
Emitter -- B
Emitter -- C

rectangle "D: Photodiode damage" as D
rectangle "E: Amplifier malfunction" as E
rectangle "F: Electrical connection issues" as F
Detector -- D
Detector -- E
Detector -- F

rectangle "G: Motor breakdown" as G
rectangle "H: Mirror/prism damage" as H
rectangle "I: Bearing wear" as I
Scanner -- G
Scanner -- H
Scanner -- I

rectangle "J: Algorithm bugs" as J
rectangle "K: Insufficient processing power" as K
Processing -- J
Processing -- K

rectangle "L: Adverse weather" as L
rectangle "M: Direct sunlight glare" as M
rectangle "N: Reflective surfaces" as N
Interference -- L
Interference -- M
Interference -- N

rectangle "O: Debris accumulation" as O
rectangle "P: Impact damage" as P
rectangle "Q: Misalignment from vibration" as Q
Obstruction -- O
Obstruction -- P
Obstruction -- Q

note bottom of TopEvent
  All connections represent OR gates
  Any lower-level event can cause
  the higher-level failure
end note

@enduml

\end{lstlisting}

Explanation given by Perplexity in words:
Top Event:
\textit{The diagram starts with the top event "LIDAR Sensor Failure in Level 4 Autonomy". This represents the overall system failure we're analyzing.
Main Categories:
The top event is connected to four main categories of failure through an OR gate, meaning that any of these categories can lead to the top event:
a) Hardware Failure
b) Software Failure
c) Environmental Factors
d) Integration Issues
Hardware Failure:
This category includes physical component failures, such as:
Laser emitter degradation
Detector malfunction
Scanning mechanism failure
Power supply issues
Software Failure:
This covers issues related to data processing and algorithms:
Algorithm bugs
Insufficient processing power
Object detection/classification errors
Prediction algorithm failure
Environmental Factors:
These are external conditions that can affect LIDAR performance:
Adverse weather (rain, snow, fog)
Direct sunlight glare
Reflective surfaces
Debris accumulation
Integration Issues:
This category addresses problems with how the LIDAR system interacts with other vehicle systems:
Sensor fusion errors
Calibration drift
Data throughput limitations
V2X communication failure
Logical Relationships:
All connections in this diagram use OR gates, indicating that any single failure mode within a category can potentially lead to the higher-level failure.
Level 4 Autonomy Context:
A note in the diagram emphasizes that Level 4 autonomy requires enhanced sensor suites and processing capabilities to handle complex urban environments and all-weather conditions. This reflects the increased demands on LIDAR systems in highly autonomous vehicles.}

Based on the explanation given by Perplexity, one can conclude that by providing just a high level statement, the LLM is capable to break down the user's requirements and perform complete top-down analysis. This shows that on the technical front , the coverage is good. We now have to work on the Representation front. For this, we first start by correcting the errors generated while previewing the PlantUML model.

\textbf{Error1}:
Syntax Error? (@ Diagram Line 26, File Line 27)
"HardwareOR -- rectangle "Detector malfunction""
(Error)

We wanted to promt engineer Perplexity to debug errors. So we entered the errors as it is in the prompts. Following were the set of error prompts which Perplexity went on learning to modify its code to get a proper diagram:\\
1. there is some issue with HardwareOR -- HardwareAND\\
2. getting error for above plantuml code: Syntax Error? (@ Diagram Line 34, File Line 35)\\
"HardwareOR -- rectangle "Detector malfunction""
(Error)\\
3. Syntax Error? (@ Diagram Line 25, File Line 26)
"HardwareAND -down-> rectangle "Power supply issues""
(Error)\\
4. getting error: Syntax Error? (@ Diagram Line 5, File Line 6)
" arc -90 to (0,-10)"
(Error)\\
5. Syntax Error? (@ Diagram Line 5, File Line 6)
" arc"
(Error)\\
6. Syntax Error? (@ Diagram Line 12, File Line 13)
%AND_GATE(HardwareAND)
\text{AND\_GATE(HardwareAND)}
(Error)\\
7. Syntax Error? (@ Diagram Line 9, File Line 10)
"diamond HardwareAND"
(Error)

We used the code generated by Perplexity, not shown in our paper currently as its for seven sets of errors and erroneous , in our VS code environment and copied the errors received in the above sequence into Perplexity.

\subsubsection{Final generated PlantUML code and diagram}
\ref{lst:PlantUMLFinalFTA} is the code snippet given by Perplexity after training it with different error prompts.

\begin{lstlisting}
[caption{\textbf{Code to generate Perplexity generated FTA after resolution of all errors using PlantUML}}, label=lst:PlantUMLFinalFTA]
@startuml LIDAR Sensor Failure FTA

skinparam rectangle {
    roundCorner 25
}

rectangle "LIDAR Sensor Failure" as TopEvent

circle MainOR
circle HardwareAND
circle SoftwareOR
circle EnvironmentalAND

rectangle "Hardware Failure" as HardwareFailure
rectangle "Software Failure" as SoftwareFailure
rectangle "Environmental Factors" as EnvironmentalFactors

TopEvent -down-> MainOR : OR

MainOR -down-> HardwareFailure
MainOR -down-> SoftwareFailure
MainOR -down-> EnvironmentalFactors

HardwareFailure -down-> HardwareAND : AND
HardwareAND -down-> (Laser emitter degradation)
HardwareAND -down-> (Power supply issues)

SoftwareFailure -down-> SoftwareOR : OR
SoftwareOR -down-> (Algorithm bugs)
SoftwareOR -down-> (Insufficient processing power)

EnvironmentalFactors -down-> EnvironmentalAND : AND
EnvironmentalAND -down-> (Adverse weather)
EnvironmentalAND -down-> (Direct sunlight glare)

note right of TopEvent
  Circles labeled 'OR' represent OR gates
  Circles labeled 'AND' represent AND gates
end note

@enduml
\end{lstlisting}

\begin{table*}[htbp]
\centering
\begin{tabular}{c|c}
\hline
\multicolumn{2}{c}{\includegraphics[width=\linewidth,height=3.0in]{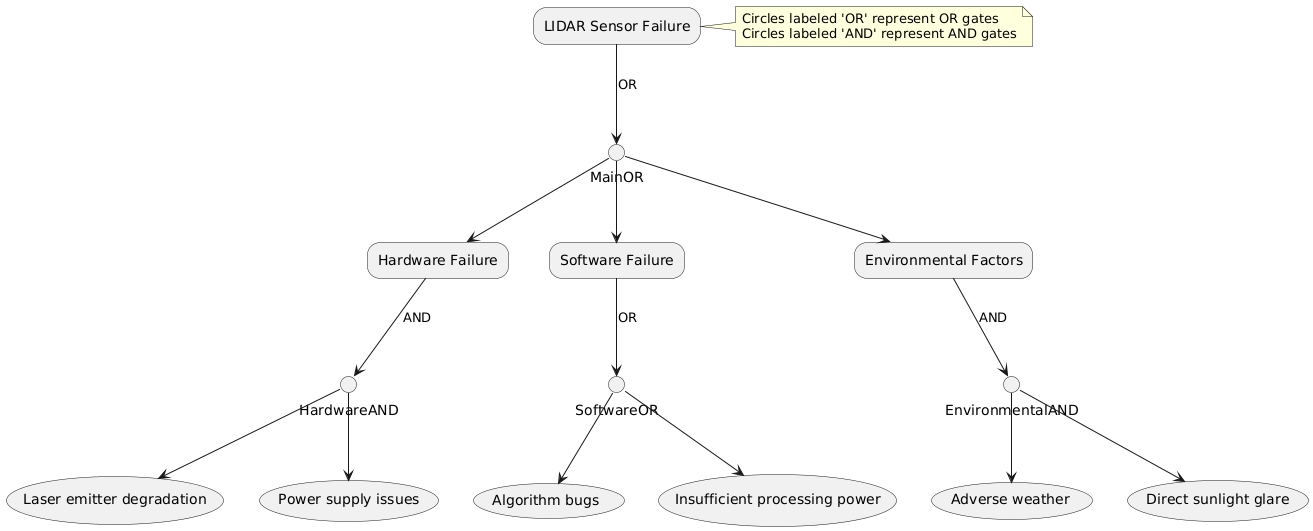}}\\
\hline
\end{tabular}
\caption{Plant UML generated FTA after resolution of all errors}
\label{Plant UML generated FTA after resolution of all errors}
\end{table*}

Code snippet\ref{lst:PlantUMLFinalFTA} and the figure\ref{Plant UML generated FTA after resolution of all errors} are the final code blocks and FTA respectively which meet our initial requirement to get a proper breakdown of all critical events necessary to perform Fault analysis of a Lidar Sensor Failure with appropriate GATES, simplified and representation as per Industry standards.

\subsection{PlantUML FTA for additional events}
We tried to further enhance the \ref{lst:PlantUMLperformanceFTA} model by adding additional possible events for Lidar failure. It understood the task and generated the UML code for the PlantUML model \ref{lst:PlantUMLperformanceFTA} shown below:
%Performance degradation code

\begin{lstlisting}
[caption{\textbf{Code to generate Perplexity generated FTA after adding additional performance degradation factor using PlantUML}}, label=lst:PlantUMLperformanceFTA]
@startuml LIDAR Sensor Failure or Degradation in Level 4 Autonomy

skinparam rectangle {
    roundCorner 25
}

rectangle "LIDAR Sensor Failure or Significant Degradation\nin Level 4 Autonomy" as TopEvent
rectangle "Hardware Failure" as HardwareFailure
rectangle "Software Failure" as SoftwareFailure
rectangle "Environmental Factors" as EnvironmentalFactors
rectangle "Integration Issues" as IntegrationIssues
rectangle "Performance Degradation" as PerformanceDegradation

TopEvent --> HardwareFailure
TopEvent --> SoftwareFailure
TopEvent --> EnvironmentalFactors
TopEvent --> IntegrationIssues
TopEvent --> PerformanceDegradation

HardwareFailure --> "Laser emitter degradation"
HardwareFailure --> "Detector malfunction"
HardwareFailure --> "Scanning mechanism failure"
HardwareFailure --> "Power supply issues"

SoftwareFailure --> "Algorithm bugs"
SoftwareFailure --> "Insufficient processing power"
SoftwareFailure --> "Object detection/classification errors"
SoftwareFailure --> "Prediction algorithm failure"

EnvironmentalFactors --> "Adverse weather (rain, snow, fog)"
EnvironmentalFactors --> "Direct sunlight glare"
EnvironmentalFactors --> "Reflective surfaces"
EnvironmentalFactors --> "Debris accumulation"

IntegrationIssues --> "Sensor fusion errors"
IntegrationIssues --> "Calibration drift"
IntegrationIssues --> "Data throughput limitations"
IntegrationIssues --> "V2X communication failure"

PerformanceDegradation --> "Range reduction"
PerformanceDegradation --> "Resolution decrease"
PerformanceDegradation --> "Increased noise in point cloud"
PerformanceDegradation --> "Slower scan rate"

note right of TopEvent
  Level 4 autonomy requires enhanced
  sensor suites and processing capabilities
  to handle complex urban environments
  and all-weather conditions. Some failures
  require multiple conditions (AND gates).
end note

@enduml
\end{lstlisting}

%\begin{table*}[htbp]
%\centering
%\begin{tabular}{c|c}
%\hline
%\multicolumn{2}{c}%{\includegraphics[width=\linewidth,height=2.5in]{perform_degradation_old.PNG}}\\
%\hline
%\end{tabular}
%\caption{Plant UML generated FTA with performance degradation added to events}
%\label{Plant UML generated FTA with performance degradation added to events}
%\end{table*}

\begin{figure*}[htpb]
    \centering
    \begin{turn}{90}
    \includegraphics[width=1.2\textwidth]{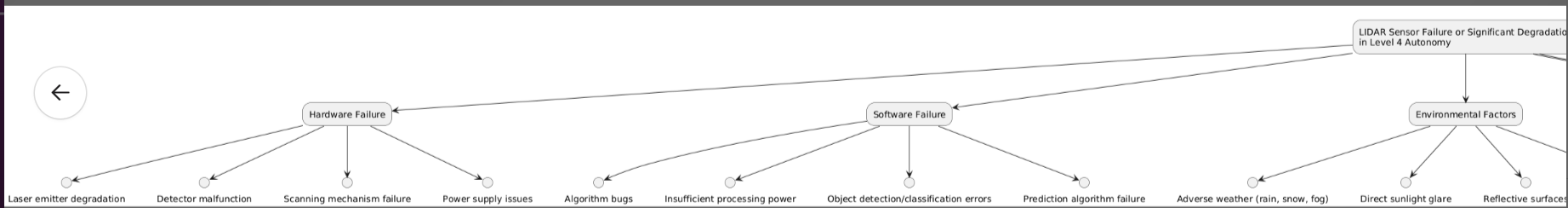}
    \end{turn}
    \caption{Plant UML generated FTA with performance degradation added to events left part of flowchart}
    \label{fig:Plant UML generated FTA with performance degradation added to events left part of flowchart}
\end{figure*}

\begin{figure*}[htpb]
    \centering
    \begin{turn}{90}
    \includegraphics[width=1.2\textwidth]{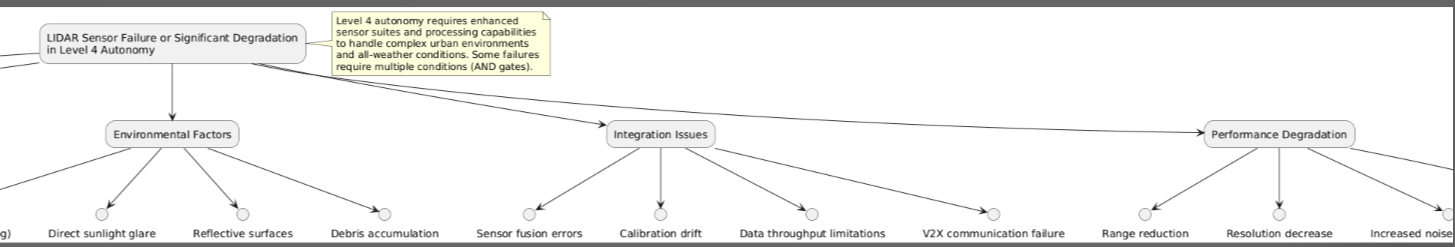}
    \end{turn}
    \caption{Plant UML generated FTA with performance degradation added to events right part of flowchart}
    \label{fig:Plant UML generated FTA with performance degradation added to events right part of flowchart}
\end{figure*}

Code snippet\ref{lst:PlantUMLperformanceFTA} and the figure\ref{fig:Plant UML generated FTA with performance degradation added to events left part of flowchart} and \ref{fig:Plant UML generated FTA with performance degradation added to events right part of flowchart} are the final code blocks and FTA diagram respectively which meet our requirements to get an additional critical event added to perform Fault analysis of a Lidar Sensor Failure. 
These have been split into two diagrams for better viewability as the diagrams were getting blurry when included in paper as a whole.

\section{Observations}

1. Initially the model was not using standard FTA shapes. When prompted to use AND/OR gates, the model ran into errors as PlantUML previewer could not understand complicated FTA shapes. Upon asking Perplexity itself to resolve errors, the model kind of back traced its errors to find an intermediate solution which used standard shapes and performs FTA.

2. The model performance did not degrade and it provided the necessary UML code when we added additional events for FTA cause.

3. The model understands the usage of AND/OR gates.

\section{Conclusion and Future Work}
Based on the learnings from training the LLM model, we conclude that the Perplexity model very well understands the technical aspects on the derivation of FTA for a particular failure. The model however needs sufficient training to get it in the right format with gates and ensure correctness as it modifies the UML code. As a next step, we will be using this for FMEA analysis in our next paper.
Current paper majorly focuses on one particular usecase for Lidar failure. There is scope for lot of work to improvise on this by using various FTA usecases for sensors like cameras, Radars and their respective faults in the autonomous vehicle field. For our analysis, we focused on Perplexity and other open source LLM models. As a future scope , one can generate an LLM model specifically which does FTA for Autonomous features and the research in this paper can serve as ground work for the same.

Also, once the new LLM model has been trained to output the entire FTA diagram with the relevant logic representation and blocks, PlantUML would be completely replaced with the LLM itself. This should enable performing the entire FTA using LLM with a single or fewer prompts. This capability of FTA generation from the LLM could then be used in different applications and use cases such as healthcare, aerospace, and others.

%\section*{APPENDIX}
%Appendixes, if needed, appear before the acknowledgment.

%\section*{ACKNOWLEDGMENT}

%\section*{REFERENCES}

\bibliographystyle{unsrt}
\bibliography{Final_trans}

\begin{thebibliography}{10}

\bibitem{KAPITONOV2022556}
Mikhail~V. Kapitonov.
\newblock Evaluation and analysis of risks in automotive industry.
\newblock {\em Transportation Research Procedia}, 61:556--560, 2022.
\newblock XII International Conference on Transport Infrastructure: Territory Development and Sustainability (TITDS-XII).

\bibitem{pathak2024iso}
Isha Pathak and Bhushan Kothari.
\newblock Iso 26262 functional safety--an approach for compliance readiness.
\newblock Technical report, SAE Technical Paper, 2024.

\bibitem{schnellbach2019development}
Adam Schnellbach and Gerhard Griessnig.
\newblock Development of the iso 21448.
\newblock In {\em Systems, Software and Services Process Improvement: 26th European Conference, EuroSPI 2019, Edinburgh, UK, September 18--20, 2019, Proceedings 26}, pages 585--593. Springer, 2019.

\bibitem{automated2020avsc}
Automated Vehicle~Safety Consortium et~al.
\newblock Avsc best practice for describing an operational design domain: Conceptual framework and lexicon.
\newblock {\em SAE Industry Technologies Consortia}, 2020.

\bibitem{10463265}
Mofeoluwa Jide-Jegede and Tomiwa Omotesho.
\newblock Harnessing generative ai for manufacturing innovation: Applications and opportunities.
\newblock In {\em 2024 International Conference on Artificial Intelligence in Information and Communication (ICAIIC)}, pages 568--572, 2024.

\bibitem{gurney2018introduction}
Kevin Gurney.
\newblock {\em An introduction to neural networks}.
\newblock CRC press, 2018.

\bibitem{rojas1996backpropagation}
Raul Rojas and Ra{\'u}l Rojas.
\newblock The backpropagation algorithm.
\newblock {\em Neural networks: a systematic introduction}, pages 149--182, 1996.

\bibitem{nyarukowa2016swapdt}
Christopher Nyarukowa, Robert Koech, Theodor Loots, and Zeno Apostolides.
\newblock Swapdt: a method for short-time withering assessment of probability for drought tolerance in camellia sinensis validated by targeted metabolomics.
\newblock {\em Journal of plant physiology}, 198:39--48, 2016.

\bibitem{simhambhatla2019self}
Ramesh Simhambhatla, Kevin Okiah, Shravan Kuchkula, and Robert Slater.
\newblock Self-driving cars: Evaluation of deep learning techniques for object detection in different driving conditions.
\newblock {\em SMU Data Science Review}, 2(1):23, 2019.

\bibitem{chowdhary2020natural}
KR1442 Chowdhary and KR~Chowdhary.
\newblock Natural language processing.
\newblock {\em Fundamentals of artificial intelligence}, pages 603--649, 2020.

\bibitem{10124508}
Temitope~Alaba Oshin and Ayomide~Josephine Tawose.
\newblock Predictive risk assessment of heart failure using hazop and qualitative risk analyses.
\newblock In {\em 2023 International Conference on Science, Engineering and Business for Sustainable Development Goals (SEB-SDG)}, volume~1, pages 1--8, 2023.

\bibitem{European_Standards}
2019 Oct.~13, 2024.~[Online].
\newblock European standards, https://www.en-standard.eu/new-aiag-vda-fmea-handbook-failure-mode-and-effects-analysis/?, 2019.

\bibitem{HarnessingAI}
Joel Thurlby.
\newblock Harnessingai, 2023.

\bibitem{basnet2018review}
Sunil Basnet, Osiris~Alejandro Valdez~Banda, Pentti Kujala, et~al.
\newblock Review of the safety engineering techniques for a complex ship system.
\newblock In {\em The 7th Asia Conference on Earthquake Engineering}. ACEE, 2018.

\bibitem{campean2008function}
Felician Campean and Ed~Henshall.
\newblock A function failure approach to fault tree analysis for automotive systems.
\newblock {\em SAE International}, pages 01--0846, 2008.

\bibitem{swarup2014safety}
M~Ben Swarup and K~Amaravathi.
\newblock Safety-critical failure analysis of industrial automotive airbag system using fmea and fta techniques.
\newblock {\em International Journal of Advanced Research in Computer Science}, 5(5), 2014.

\bibitem{silvis2024sart}
Natalia Silvis-Cividjian, Yijing Zhou, Anastasia Sarchosoglou, and Evangelos Pappas.
\newblock i-sart: An intelligent assistant for safety analysis in radiation therapy.
\newblock In {\em BIOSTEC (2)}, pages 420--427, 2024.

\bibitem{gimpel2023unlocking}
Henner Gimpel, Kristina Hall, Stefan Decker, Torsten Eymann, Luis L{\"a}mmermann, Alexander M{\"a}dche, Maximilian R{\"o}glinger, Caroline Ruiner, Manfred Schoch, Mareike Schoop, et~al.
\newblock Unlocking the power of generative ai models and systems such as gpt-4 and chatgpt for higher education: A guide for students and lecturers.
\newblock Technical report, Hohenheim Discussion Papers in Business, Economics and Social Sciences, 2023.

\bibitem{lu2022safety}
Jiayi Lu, Shichun Yang, Runwu Shi, Yuyi Chen, and Yaoguang Cao.
\newblock Safety co-pilot: A system for autonomous vehicle to make decision safer and smarter.
\newblock In {\em 2022 IEEE 25th International Conference on Intelligent Transportation Systems (ITSC)}, pages 1809--1816. IEEE, 2022.

\bibitem{backstrom2016effective}
Ola B{\"a}ckstr{\"o}m, Yuliya Butkova, Holger Hermanns, Jan Kr{\v{c}}{\'a}l, and Pavel Kr{\v{c}}{\'a}l.
\newblock Effective static and dynamic fault tree analysis.
\newblock In {\em Computer Safety, Reliability, and Security: 35th International Conference, SAFECOMP 2016, Trondheim, Norway, September 21-23, 2016, Proceedings 35}, pages 266--280. Springer, 2016.

\bibitem{tankaas}
Single-shell Tank.
\newblock Aas 00 seeettm data transmittal page.
\newblock {\em NA}, NA.

\bibitem{leveson2014comparison}
Nancy Leveson, Chris Wilkinson, Cody Fleming, John Thomas, and Ian Tracy.
\newblock A comparison of stpa and the arp 4761 safety assessment process.
\newblock {\em Massachusetts Institute of Technology, Cambridge, MA}, 2014.

\bibitem{moon1966use}
DONALD MOON, DONALD SHINN, and WALTER HYLER.
\newblock Use of statistical considerations in establishing design allowables for military handbook 5.
\newblock In {\em Symposium on Deep Submergence Propulsion and Marine Systems}, page 2522, 1966.

\bibitem{zornoza2018model}
Enrique Zornoza~Moreno.
\newblock Model-based approach for automatic generation of iec-61025 standard compliant fault trees, 2018.

\bibitem{FTA_Report}
United State Dept Of~Transportation Aug~9, 2024.~[Online].
\newblock Fta report, https://www.transit.dot.gov/research-innovation/mobility-data-standards-and-specifications-interoperability-report-0267.

\bibitem{SixSigma}
Ted~Hessing Oct.~13, 2024.~[Online].
\newblock Sixsigma fta, https://sixsigmastudyguide.com/fault-tree-analysis/, 2020.

\bibitem{ChatGPT}
OpenAI Oct.~13, 2024.~[Online].
\newblock Chatgpt, https://chatgpt.com/.

\bibitem{Gemini}
GOOGLE Oct.~13, 2024.~[Online].
\newblock Gemini, https://gemini.google.com/.

\bibitem{Perplexity}
Denis~Yarats Oct.~13, 2024. [Online] Aravind Srinivas Andy~Konwinski and Johnny Ho.
\newblock Perplexity, https://www.perplexity.ai/.

\bibitem{bing}
Bing.
\newblock bing video, https://www.bing.com/videos/riverview/relatedvideo?\\q=research+links+to+do+fta+nalysis\\`i\&'mid=1c7f6042d569d7fdd4af1c7f6042d569d7fdd4af`i\&'form \\=vire.

\bibitem{wiley}
2024. [Online] John Wiley `I\&' Sons~Inc. Sept~5.
\newblock wiley, https://onlinelibrary.wiley.com/doi/full/10.1002/qre.3271.

\bibitem{lund2022towards}
Jonas Lund, Lucas~Bjarke Jensen, Hugo~Daniel Macedo, and Peter~Gorm Larsen.
\newblock Towards uml and vdm support in the vs code environment.
\newblock In {\em Proceedings of the 20th International Overture Workshop}, pages 50--65, 2022.

\bibitem{HowToUML}
CS~Odessa Oct.~25, 2024.~[Online].
\newblock Howtouml, https://www.conceptdraw.com/how-to-guide/diagrams-software-tools-for-design-uml-collaboration-diagram.

\bibitem{FTA_Tool}
Reliotech Oct~15, 2024.~[Online].
\newblock Fta tool,https://www.fault-tree-analysis.com/.

\bibitem{Mermaid_Tool}
Gabriel~Valiente Oct~15, 2024.~[Online].
\newblock Mermaid tool, https://mermaid.live/.

\bibitem{yUML_Tool}
Various~Contributors Oct~15, 2024.~[Online].
\newblock yuml tool, https://yuml.me/.

\bibitem{nomnoml_Tool}
Various~Contributors Oct~15, 2024.~[Online].
\newblock nomnoml tool, https://nomnoml.com/.

\bibitem{kroki_Tool}
Various~Contributors Oct~15, 2024.~[Online].
\newblock kroki tool, https://kroki.io/.

\end{thebibliography}

%\bibitem{b11}
%Sajad Ebrahimi, Kimberly Vachal, Joseph Szmerekovsky, A Delphi-FMEA model to assess county-level speeding crash risk in North Dakota, Transportation Research Interdisciplinary Perspectives, Volume 16, 2022, 100688,ISSN 2590-1982,

\begin{IEEEbiography}[{\includegraphics[width=1in,height=1.25in,clip,keepaspectratio]{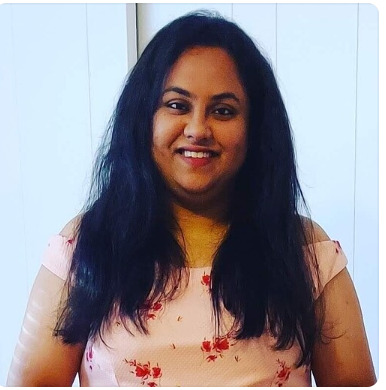}}]{Sneha Sudhir Shetiya} is a Senior IEEE Member and received her Maters degree in Electrical Engineering with a major in computer vision and Signal Processing from North Carolina State University, in 2021. She received her Bachelor's degree in Electronics and Communication Engineering from the Visvesvaraya Technological University, Karnataka, India, in 2014. Her work involves middleware topics for embedded development of autonomous driving stack, automotive diagnostics, systems engineering and functional safety. She is an active volunteer with IEEE region 4 and takes part in activities of Women In Engineering (WIE) groups in the region. She is part of the committee for senior member evaluation at IEEE for 2024 and has been a proctor for IEEExtreme 24 hour coding competition.
\end{IEEEbiography}
\begin{IEEEbiography}[{\includegraphics[width=1in,height=1.25in,clip,keepaspectratio]{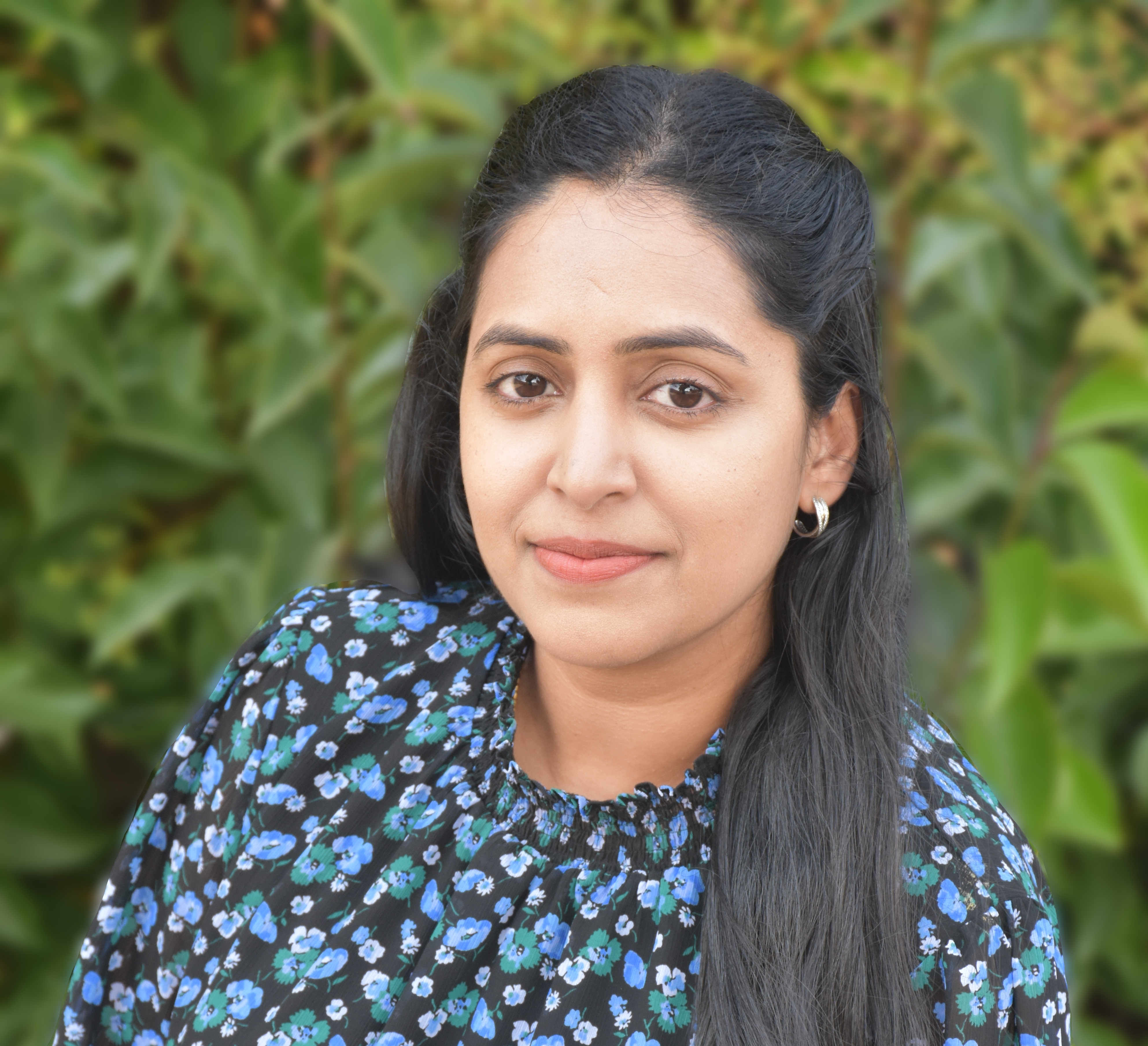}}]{Divya Garikapati} is a safety expert and currently serving as the standards committee member within the IEEE Intelligent Transportation Systems Society (ITSS) and a peer reviewer for IEEE Intelligent Transportation Systems Conferences (ITSC). She actively participates in several industry level standards discussions within IEEE and SAE organizations. She is also the working group chair for the IEEE Vehicular Technology Society standards discussions. She received her Masters in Electrical Engineering Systems from the University of Michigan, Ann Arbor in 2014. Prior to that, she received her Bachelors in Electronics and Communications Engineering from Andhra University College of Engineering, Andhra Pradesh, India. Her current work focuses on Systems and Safety research and development for Level 2,3 and 4 Autonomous vehicles. She has over 10 years of experience in the automotive industry.
\end{IEEEbiography}
\begin{IEEEbiography}[{\includegraphics[width=1in,height=1.25in,clip,keepaspectratio]{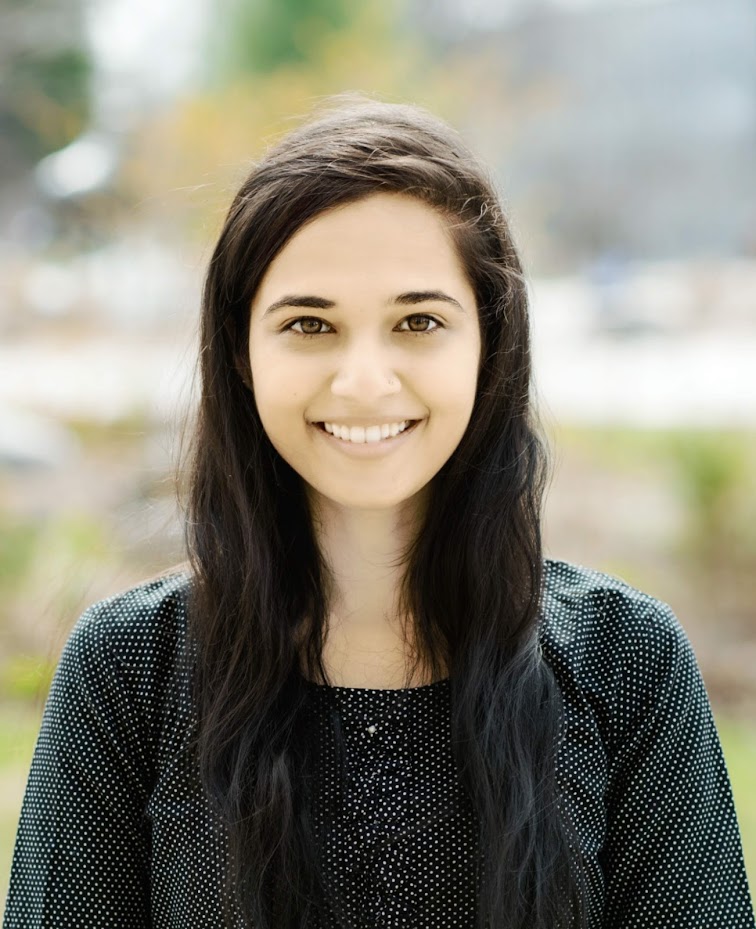}}]{Veeraja Sohoni} is a Senior Engineer working in Automotive safety with a focus on functional safety. She has 7 years of experience in Automotive Safety working at renowned Organizations. She graduated from Michigan Technological University in May 2021 with a Masters in Electrical Engineering. She was a CHEB member of Society of Women Engineers at Michigan Tech. Her work involves supporting stakeholder requirement elicitation for SoC, its analysis and modelling various simulated use-cases in closed loop.
\end{IEEEbiography}

\end{document}